\documentclass[aps,prb,groupedaddress,twocolumn,showpacs,preprintnumbers,amsmath,amssymb]{revtex4}

\usepackage{graphicx}
\usepackage{dcolumn}
\usepackage{bm}
\usepackage{eucal} 
\usepackage{ulem}
\usepackage{color}
\usepackage{afterpage}
\newcommand{\cacoo}{Ca$_3$Co$_2$O$_6$}
\begin{document}

\title{Magnetic nano-fluctuations in a frustrated magnet}
\author{Krunoslav~Pr\v{s}a,$^{1,2,3}$ Mark~Laver,$^4$ Martin~M{\aa}nsson,$^1$ Sebastian~Guerrero,$^5$ Peter~M.~Derlet,$^5$ 
Ivica~\v{Z}ivkovi{\'c},$^6$ Hee~Taek~Yi,$^7$ Lionel~Porcar,$^8$ Oksana~Zaharko,$^1$ Sandor~Balog,$^9$ Jorge~L.~Gavilano,$^1$
Joachim~Kohlbrecher,$^1$ Bertrand~Roessli,$^1$ Christof~Niedermayer,$^1$ Jun~Sugiyama,$^{10}$ Cecile~Garcia,$^{11}$ Henrik~M.~R{\o}nnow,$^3$
Christopher~Mudry,{$^5$} Michel~Kenzelmann,$^{12}$ Sang--Wook~Cheong$^7$ and Jo\"{e}l~Mesot$^{1,2,3}$}
\affiliation{
$^1$Laboratory for Neutron Scattering, Paul Scherrer Institute, CH-5232 Villigen PSI, Switzerland\\
$^2$Laboratory for Solid state physics, ETH Z\"urich, Switzerland\\
$^3$Laboratory for Quantum magnetism, EPFL, CH-1015 Lausanne, Switzerland\\
$^4$School of Metallurgy and Materials, University of Birmingham, Birmingham B15 2TT, UK\\
$^5$Condensed Matter Theory Group, Paul Scherrer Institute, CH-5232 Villigen PSI, Switzerland\\
$^6$Institute of Physics, Bijeni\v{c}ka 46, HR-10000 Zagreb, Croatia\\
$^7$Rutgers Center for Emergent Materials and Department of Physics \& Astronomy, Rutgers University, Piscataway, New Jersey 08854, USA\\
$^8$Institut Laue Langevin, 6 rue Jules Horowitz, 38042 Grenoble Cedex 9, France\\
$^9$Adolphe Merkle Institute, University of Fribourg, Rte de l'Ancienne Papeterie, PO Box 209, CH-1723 Marly 1, Switzerland\\
$^{10}$Toyota Central Research and Development Laboratories, Inc., Nagakute, Aichi 480-1192, Japan\\
$^{11}$Universit\'e de Toulouse, INSA-LPCNO, CNRS UMR 5215, 135 Av.\ de Rangueil, F-31077 Toulouse, France\\
$^{12}$Laboratory for Developments and Methods, Paul Scherrer Institute, CH-5232 Villigen PSI, Switzerland\\
}

\date{\today}
\maketitle

\textbf{Frustrated systems exhibit remarkable properties due to the high degeneracy of their ground states.
Stabilised by competing interactions\cite{Seul1995}, a rich diversity of typically nanometre--sized phase structures appear in polymer\cite{Seo2012}
and colloidal\cite{Adams1998} systems, while the surface of ice pre-melts\cite{Watkins2011} due to geometrically frustrated interactions.
Atomic spin systems where magnetic interactions are frustrated by lattice geometry provide a fruitful source of
emergent phenomena, such as fractionalised excitations analogous to magnetic monopoles\cite{Ladak2010,Castelnovo2012}.
The degeneracy inherent in frustrated systems may prevail all the way down to absolute zero temperature\cite{Balents2010},
or it may be lifted by small perturbations\cite{Shokef2011} or entropic effects~\cite{Savary2012}. In the geometrically frustrated Ising--like magnet \cacoo,
we follow the temporal and spatial evolution of nanoscale magnetic fluctuations firmly embedded inside the spin--density--wave magnetic structure.
These fluctuations are a signature of a competing ferrimagnetic phase with an incommensurability that is different from, but determined by the host.
As the temperature is lowered, the fluctuations slow down into a super-paramagnetic regime of stable spatiotemporal nano-structures.}

\begin{figure*}
\includegraphics[width=6.4in]{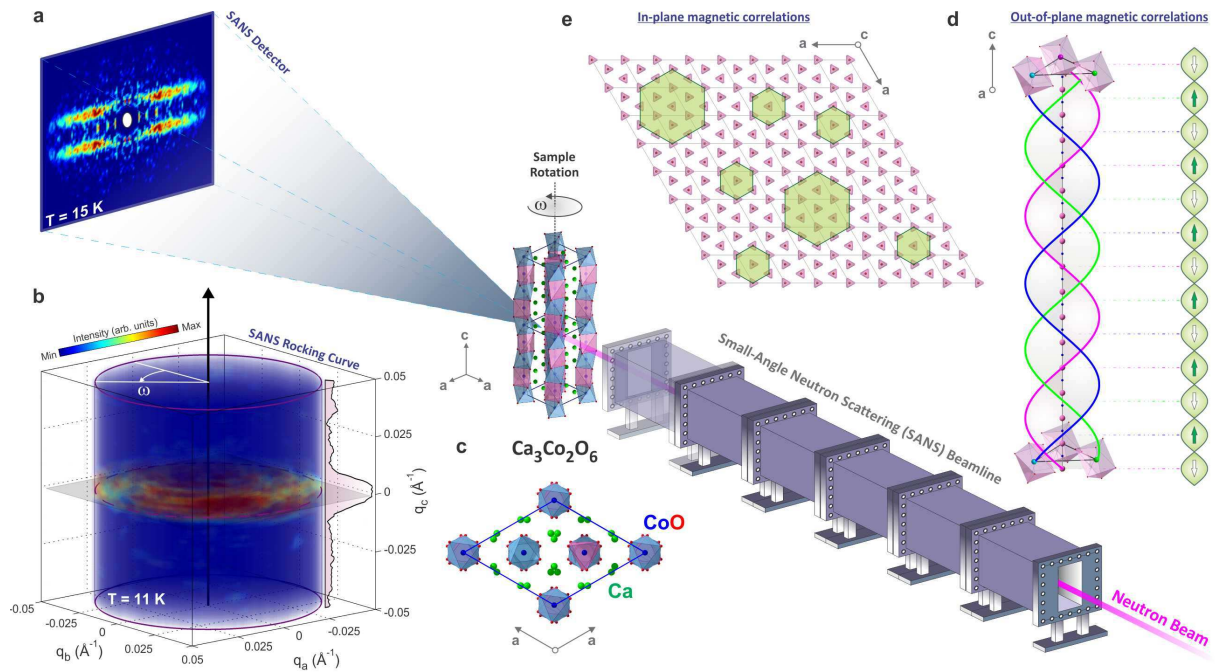}
\caption{\textbf{Nano-fluctuations imaged by small--angle neutron scattering in \cacoo.} \textbf{a}, The large central schematic illustrates the experimental setup and scattering observed on the small--angle multidetector from the nano-fluctuations embedded in the modulated PDA background. For $T < T_\mathrm{N} \simeq 25$\,K, streaks are observed when the $c$-axis of the single crystal \cacoo\ is in the plane of the detector. \textbf{b}, The scattering volume in three dimensions measured by rotating a single crystal about the $c$-axis reveals isotropic scattering in the $ab$~plane. The sample region in the centre of the figure shows the arrangement of magnetic Co$^{3+}$ ions in \cacoo, which lie in chains along the $c$-axis made up of face-sharing CoO$_6$ polyhedra alternating between {blue} low-spin ($S=0$) octahedra and {red} high-spin ($S=2$) trigonal prisms\cite{Aasland1997}. \textbf{c}, A projection along the $c$-axis showing that the chains form a triangular lattice in the $ab$~plane. \textbf{d}, The magnetic structure along the $c$-axis consists of phase-shifted spin-density waves. We propose that the objects seen in SANS originate from areas of enhanced PDA fluctuations originating at nodal points of the three spin-density waves. \textbf{e}, In the $ab$--plane these nano-fluctuations are ferrimagnetic and have dimensions spanning several unit cells.} 
\end{figure*}

The Ising model on a two-dimensional (2D) triangular lattice with antiferromagnetic nearest-neighbour interactions $J_\mathrm{ab} > 0$
has long embodied the {archetype} of a geometrically frustrated spin system\cite{Wannier1950,Houtappel1950}.
It is convenient to divide the triangular lattice into three sublattices, whereby every elementary triangle contains one site of each sublattice connected by three antiferromagnetic bonds that cannot be simultaneously satisfied. One bond per triangle remains frustrated at the minimum energy. The extensive ground-state degeneracy, equal to $\exp (0.323 N)$ as the number of spins $N \to \infty$,
means that the system is a spin liquid at finite temperatures above the $T=0$ critical point\cite{Wannier1950,Houtappel1950}.
The residual entropy can be lifted by increasing the dimensionality through the stacking of layers of triangular lattices on top of each other.
Since the interaction~$J_c$ along the stacking direction~$c$ is unfrustrated, the ground state degeneracy reduces to $\exp (0.323 N^{2/3})$.
The result is a transition at $T > 0$ to a partially disordered antiferromagnetic (PDA) state. This state has long-range order along~$c$ but only two out of three sublattices are ordered within the $ab$ plane, while the third fluctuates.
The competing ferrimagnetic state, with the third sublattice direction fixed, has equal energy to the PDA, but the PDA is favoured for entropic reasons\cite{Jiang2006}.

{Small angle neutron scattering (SANS) experiments were performed on \cacoo\ to investigate the small $\mathbf{Q}$ reciprocal space structure (Fig.1, central schematic and a).} In the frustrated magnet \cacoo, the magnetic Co$^{3+}$ ions form structural chains along the $c$-axis, with a buckled triangular arrangement in the $ab$~plane~(Fig.~1{c}) resembling the stacked triangular model. The moments point along the chains with an Ising--like anisotropy\cite{Wu2005,Jain2013}, but below the ordering temperature at $T = T_\mathrm{N} \simeq 25$~K are amplitude--modulated by an incommensurate longitudinal spin density wave (SDW) propagating along the $c$-axis, with a phase shift of $120^\circ$ between adjacent chains\cite{Agrestini2008PRB,Agrestini2008PRL} {(Fig.~1d)}. Every one--sixth of the modulation period, a rigorous PDA condition (an ``up''--``down''--``zero'' expectation value for the sublattice magnetizations) holds, presumably accompanied by a strong fluctuation of the Ising--like spins. This modulation is thought to be stabilised by interchain interaction pathways\cite{Fresard2004,Chapon2009,Kamiya2012}.

Here, we uncover ferrimagnetic nanoscale fluctuations forming within the SDW ordered state at all $T < T_\mathrm{N}$ in \cacoo\ (Fig.~1{a}). To scatter neutrons through small angles by reciprocal vectors $\mathbf{Q}$ in the vicinity of the $\mathbf{Q}=\mathbf{0}$ position, a local ferromagnetic component of spin correlations suffices. These dynamic objects produce two parallel streaks at incommensurate positions on the 2D SANS detector (Figs.~{2a-b}) running perpendicular to the $c$-axis which is aligned in the plane of the detector. This incommensurability is connected with a third of the real-space periodicity $d/3$ of the SDW modulation along~$c$, reflecting the strong PDA--like fluctuation points in the ground state of the material (Fig.~{1c}). The scattering pattern does not change when the sample is rotated about the $c$-axis (Fig.~{1b}, also see Suppl.~Info), implying that correlations are isotropic in the $ab$ plane. Accordingly, we can separate the along-chain ($Q_\mathrm{c}$) and the isotropic in-plane ($Q_\mathrm{ab}$) components, and the scattered neutron intensity $I(\mathbf{Q})$ is then described by
\begin{equation}
I(\mathbf{Q}) \propto \delta\left(Q_\mathrm{c} \pm \frac{2 \pi}d\right) \frac{Q_\mathrm{ab}^2}{Q_\mathrm{c}^2}
\int_0^\infty   \, C_\mathrm{ab}(r) J_0(Q_\mathrm{ab} r)\,r\mathrm{d}r, \label{eq:scatt}
\end{equation}
where $C_\mathrm{ab}(r)$ is the spin-spin correlation in the $ab$-plane.

\begin{figure}
\includegraphics[width=3.2in]{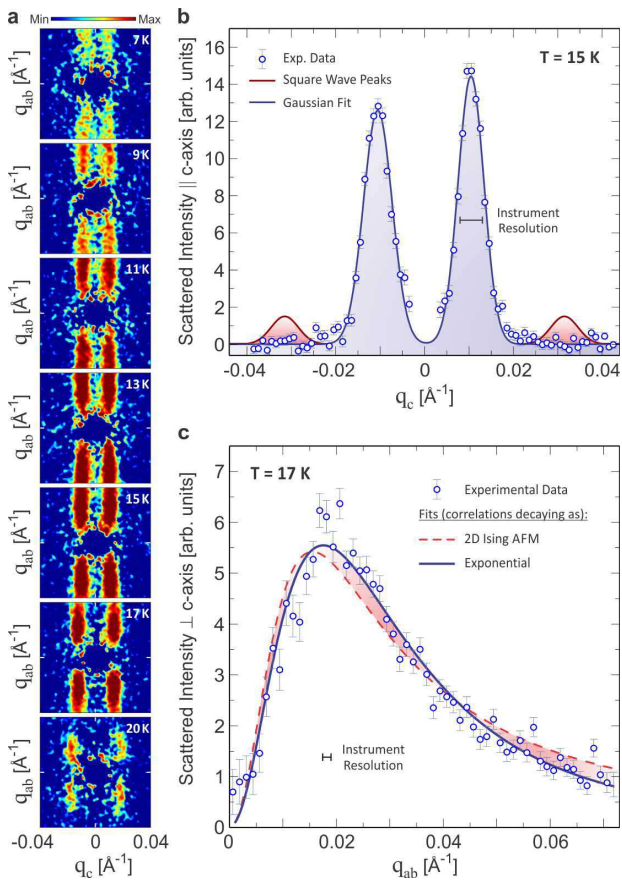}
\caption{\textbf{Characteristic scattering profiles from ferrimagnetic nanophases.} {\textbf{a}}, SANS detector images at different temperatures. {\textbf{b}}, Variation in direction normal to the streaks. The absence of higher-order peaks is demonstrative of the sinusoidal amplitude--modulation of nanofluctuations. {\textbf{c}}, Variation of scattered intensity along the streaks. This profile is a transform of the spin-spin correlation function in the triangular lattice plane, see Eq.~\eqref{eq:scatt}. }
\end{figure}

Spin correlations are further studied by looking at the $Q_\mathrm{ab}$ dependence of the scattering along the streaks (Fig.~2{c}).
From Eq.~\eqref{eq:scatt}, such profiles are Hankel transforms of the in-plane spin-spin correlation function $C_\mathrm{ab}(r)$ multiplied by $Q_\mathrm{ab}^2 / Q_\mathrm{c}^2$.
This multiplicative factor is a consequence of the dipolar interaction between neutrons and the atomic moments aligned along~$c$. In Fig.~2{c}, we show the measured profile compared to two model spin-spin correlation functions (more are compared in the Supplementary Information): (i) the finite-temperature expression for the 2D triangular lattice antiferromagnet (TLA), $C_\mathrm{ab}(r) \sim r^{-\frac12} \exp(-r/\xi_\mathrm{ab})$, where we have ignored the antiferromagnetic $\cos (2 \pi r /3)$ modulation\cite{Wojtas2009} and (ii) a purely exponentially decaying correlation, $C_\mathrm{ab}(r) \sim \exp(-r/\xi_\mathrm{ab})$. For both models $\xi_\mathrm{ab}$ is temperature dependent. The intensity scale and correlation length $\xi_\mathrm{ab}$ are considered as fit parameters to the measured data-set.
Model (i) might be anticipated to describe ferromagnetic correlations arising as perturbations of the PDA state,
and for a 2D TLA we would expect $\xi_\mathrm{ab} \sim 1/|\ln \tanh (1/T)|$ to increase as $T$ (here in units of the model's coupling constant) decreases. However, the measured $Q_\mathrm{ab}$ profiles are seen to flatten as $T$ decreases, implying a decreasing $\xi_\mathrm{ab}$.
The form of the curves of all the measured datasets (c.f.\ Supplementary Information) are best described
by the exponentially decaying correlations of model (ii). { Fitting the data of Fig.~2a to model (ii) gives the correlation length $\xi_\mathrm{ab}$ as a function of temperature (Fig.~3a) --- indeed showing a decrease of $\xi_\mathrm{ab}$ with decreasing temperature.} The in--plane exponential decay of spin-spin correlation could be consistent with ferrimagnetic regions existing at the walls between PDA
domains\cite{Matsubara1987} or a thermally induced effective interactions within the disordered sublattice (see Supplementary {Information}). The correlation length along the $c$ axis follows a trend with temperature similar to $\xi_\mathrm{ab}$  (Fig.~3a) but is longer due to the stronger interactions along the chains. At 15\,K, this structure is extended $\sim5$ interatomic distances in the $ab$ direction and $\sim90$ interatomic distances along the $c$ direction in real space (Fig.~1d).

\begin{figure}
\includegraphics[width=3.2in]{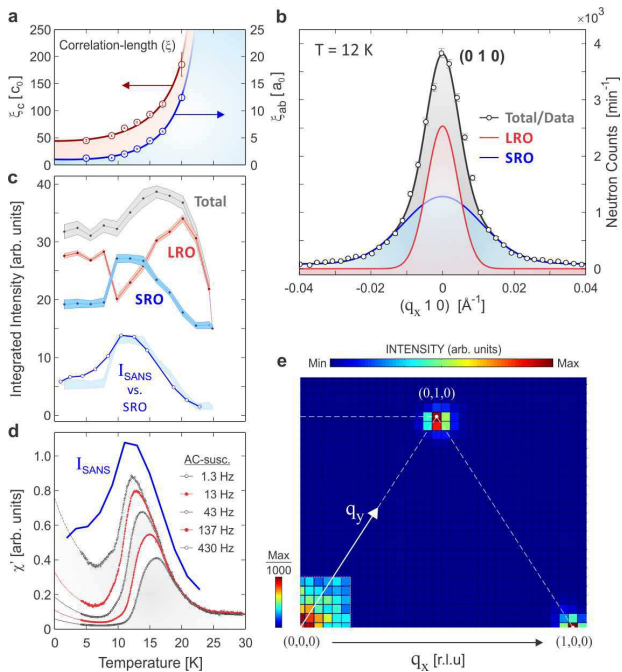}
\caption{\textbf{Temperature dependent properties of ferrimagnetic nanophases.}
\textbf{a}, Correlation lengths of SANS scattering $\xi_\mathrm{ab}$ and $\xi_\mathrm{c}$. {\textbf{b}}, A typical neutron diffraction antiferromagnetic peak showing two components with different lengthscales. The sharper Gaussian comes from the antiferromagnetic order, while the broad Gaussian relates to the SANS ($\mathbf{Q}=\mathbf{0}$) scattering and so originates from the ferrimagnetic microstructures. {\textbf{c}}, Neutron diffraction intensity of long--range order and short--range correlations. The short--range order in diffraction has a comparable $T$--dependence to the SANS intensity. {\textbf{d}}, AC susceptibility data at several frequencies with superimposed SANS intensity (blue) for comparison.  \textbf{e}, Spin-spin correlation function derived by a Monte-Carlo approach. This quantity is proportional to the magnetic neutron scattering intensity.}
\end{figure}

The nanofluctuations originate from, but are distinct from the PDA background. The latter phase is well ordered over large lengthscales $> 0.5$\,$\mu$m at $T \lesssim 25$\,K\cite{Agrestini2008PRB}, producing resolution-limited {antiferromagnetic} Bragg peaks in our neutron diffraction data (Fig.~3{b}). Here the nano-fluctuations generate broad features superimposed on the sharp antiferromagnetic peaks\cite{Agrestini2008PRL} that are seen to share a common temperature dependence with the SANS signal (Fig.~3{c}). This is a signature of ferrimagnetic nano-scale objects, which would give broad anti-ferromagnetic and $\textbf{Q}=0$ peaks. The temperature dependences of the integrated intensities reveal a competition between the two components, since the embedded {nano-}structures exist at the expense of magnetic volume fraction for the host.

Microscopic ferrimagnetism in the compound also explains the experimentally observed magnetic response functions. In agreement with previous studies\cite{Hardy2004}, we find a large cusp in the linear susceptibility $\chi'$ at temperatures above 12\,K (Fig.~3{d}). The susceptibility curve resembles the SANS intensity, indicative of a common origin. The position in temperature $T_f$ of the peak shifts with the ac frequency $f$, a dependence that is characterised by the quantity $g = \Delta T_f / [T_f \Delta (\log f)]$. Here we find $g = 0.17$, which is too big to be consistent with spin glass ($g \lesssim 0.01$) or cluster glass ($g \approx 0.05$) relaxations\cite{Mahendiran2003}. Instead larger spin structures are implied, tying in with values $g \gtrsim 0.1$ exhibited by frequency-dependent blocking transitions in superparamagnetic nanostructures\cite{Mahendiran2003}. The observed frequency dependence directly reveals the characteristic temporal scales of the nano-fluctuations. As the temperature is lowered, they slow down from the millisecond (16~K) to the second (12~K) timescale, effectively producing stable spatiotemporal objects. Large spin structures are also supported by nonlinear susceptibility measurements (see Supplementary information).

To {shed some light on} the origin of the nanophases in \cacoo, we studied the simplified system of the strongly anisotropic stacked triangular lattice. Our Monte Carlo simulations confirm a finite-temperature order--by--disorder phase transition from the disordered high-temperature phase into the partially ordered PDA. This transition is signaled by an extensive peak at the anti-ferromagnetic points in the Brillouin zone (see Fig.~3{e} and the Supplementary Information). Theoretically it is known that the ferrimagnetic state (which has the same energy as the PDA state) unsuccessfully competes with the PDA state\cite{Jiang2006,Moessner2001}. Despite this, our Monte Carlo simulations evidence residual ferrimagnetic fluctuations in the structure factor around $\mathbf{Q}=0$ which are non-extensive in their magnitude. The fact that such fluctuations do occur for our model Hamiltonian suggests that additional interactions may further stabilise ferrimagnetic fluctuations in \cacoo, leading to
a plausible explanation for the observed nanophases.
These additional interactions will likely entail a coupling of the magnetism to other degrees of freedom.
Recent dielectric measurements\cite{Basu2013} indicate magnetoelectric coupling in \cacoo, which might be envisaged to help stabilise nanophases
via local distortions of oxygen octahedra\cite{Bindu2009}, in a manner analagous to the structural nano-phase separation recently reported in the frustrated antiferromagnet
$\alpha$-NaMnO$_2$\cite{Zorko2014} or to the accommodation-strain-mediated phase separation in the colossal magnetoresistant manganites\cite{Ahn2004}.

{\it Methods Summary}
Susceptibility measurements were performed on a CryoBIND ac susceptometer. Direct measurements of the ferrimagnetic nanophases using small-angle neutron scattering were made using the SANS-I and SANS-II instruments at the Swiss Spallation Neutron Source (SINQ) and the D22 instrument at the Institut Laue Langevin. Neutron diffraction measurements on our single crystals were performed on the TASP and Rita-II triple-axis instruments at SINQ.

{\it Acknowledgements}
K.P. thanks A.~Zheludev (ETH Z\"{u}rich) for discussions. Technical assistance at PSI by M.~Bartkowiak and M.~Zolliker is highly appreciated. The work at Rutgers University was supported by the DOE under Grant No.~DE-FG02-07ER46382.


\begin{thebibliography}{01}
\bibitem{Seul1995} Seul,~M. \& Andelman,~D. Domain shapes and patterns: The phenomenology of modulated phases. \textit{Science} \textbf{267}, 476--483 (1995).
\bibitem{Seo2012} Seo,~M. \& Hillmyer,~M.~A. Reticulated nanoporous polymers by controlled polymerization-induced nanophase separation. \textit{Science} \textbf{336}, 1422--1425 (2012).
\bibitem{Adams1998} Adams,~M., Dogic,~Z., Keller,~S.~L. \& Fraden,~S. Entropically driven nanophase transitions in mixtures of colloidal rods and spheres. \textit{Nature} \textbf{393}, 349--352 (1998).
\bibitem{Watkins2011} Watkins,~M., Pan,~D., Wang,~E.~G., Michaelides,~A., Vande Vondele,~J. \& Slater,~B. Large variation of vacancy formation energies in the surface of crystalline ice. \textit{Nature Materials} \textbf{10}, 794--798 (2011).
\bibitem{Ladak2010} Ladak,~S., Read,~D.~E., Perkins,~G.~K., Cohen,~L.~F. \& Brandford,~W.~R. Direct observation of magnetic monopole defects in an artificial spin-ice system. \textit{Nature Physics} \textbf{6}, 359--363 (2010).
\bibitem{Castelnovo2012} Castelnovo,~C., Moessner,~R. \& Sondhi,~S.~L. Spin ice, fractionalization, and topological order. \textit{Annu.~Rev.~Condens.~Matter~Phys.}~\textbf{3}, 35--55 (2012).
\bibitem{Balents2010} Balents,~L. Spin liquids in frustrated magnets. \textit{Nature} \textbf{464}, 199--208 (2010).
\bibitem{Shokef2011} Shokef,~Y., Souslov,~A. \& Lubensky,~T.~C. Order by disorder in the antiferromagnetic Ising model on an elastic triangular lattice. \textit{Proc.~Nat.~Acad.~Sci.}~\textbf{108}, 11804--11809 (2011). 
\bibitem{Savary2012} Savary,~L., Ross,~K.~A., Gaulin,~B.~D., Ruff,~J.~P.~C. \& Balents,~L. Order by quantum disorder in Er$_2$Ti$_2$O$_7$. \textit{Phys.~Rev.~Lett.}~\textbf{109}, 167201 (2012).
\bibitem{Wannier1950} Wannier,~G.~H. Antiferromagnetism. The triangular Ising net. \textit{Phys.~Rev.}~\textbf{79}, 357--364 (1950).
\bibitem{Houtappel1950} Houtappel,~R.~M.~F. Order-disorder in hexagonal lattices. \textit{Physica}~\textbf{16}, 425--455 (1950).
\bibitem{Jiang2006} Jiang,~Y. \& Emig,~T. Ordering of geometrically frustrated classical and quantum triangular Ising magnets. \textit{Phys.~Rev.~B} \textbf{73}, 104452 (2006).
\bibitem{Aasland1997} Aasland,~S., Fjellv{\aa}g,~H. \& Hauback,~B. Magnetic properties of the one-dimensional \cacoo. \textit{Solid~State~Commun.} \textbf{101}, 187--192 (1997). 
\bibitem{Wu2005} Wu,~H., Haverkort,~M.~W., Hu,~Z., Khomskii,~D.~I. \& Tjeng,~L.~H. Nature of magnetism in \cacoo. \textit{Phys.~Rev.~Lett.}~\textbf{95}, 186401 (2005).
\bibitem{Jain2013} Jain,~J., Portnichenko,~P.~Y., Jang,~H., Jackeli,~G., Friemel,~G., Ivanov,~A., Piovano,~A., Yusuf,~S.~M., Keimer,~B. \& Inosov,~D.~S. One-dimensional dispersive magnon excitation in the frustrated spin-2 chain system \cacoo. \textit{Phys.~Rev.~B}~\textbf{88}, 224403 (2013).
\bibitem{Agrestini2008PRB} Agrestini,~S., Mazzoli,~C., Bombardi,~A. \& Lees,~M.~R. Incommensurate magnetic ground state revealed by resonant x-ray scattering in the frustrated spin system \cacoo. \textit{Phys.~Rev.~B}~\textbf{77}, 140403(R) (2008). 
\bibitem{Agrestini2008PRL} Agrestini,~S. \textit{et al.} Nature of the magnetic order in \cacoo. \textit{Phys.~Rev.~Lett.}~\textbf{101}, 097207 (2008). 
\bibitem{Fresard2004} Fr\'esard,~R., Laschinger,~C., Kopp,~T. \& Eyert,~V. Origin of magnetic interactions in \cacoo. \textit{Phys.~Rev.~B}~\textbf{69}, 140405(R) (2004).
\bibitem{Chapon2009} Chapon,~L.~C. Origin of the long-wavelength magnetic modulation in \cacoo. \textit{Phys.~Rev.~B}~\textbf{80}, 172405 (2009).
\bibitem{Kamiya2012}Kamiya,~Y. \& Batista,~C.~D. Formation of magnetic microphases in \cacoo. \textit{Phys.~Rev.~Lett.}~\textbf{109}, 067204 (2012).
\bibitem{Agrestini2011} Agrestini,~S. \textit{et al.} Slow magnetic order-order transition in the spin chain antiferromagnet \cacoo. \textit{Phys.~Rev.~Lett.}~\textbf{106}, 197204 (2011).
\bibitem{Hardy2004} Hardy,~V., Flahaut,~D., Lees,~M.~R. \& Petrenko,~O.~A. Magnetic quantum tunneling in \cacoo\ studied by ac susceptibility: Temperature and magnetic-field dependence of the spin-relaxation time. \textit{Phys.~Rev.~B}~\textbf{70}, 214439 (2004).
\bibitem{Wojtas2009} Wojtas,~D.~H. \& Millane,~R.~P. Two-point correlation function for the triangular Ising antiferromagnet. \textit{Phys.~Rev.~B}~\textbf{79}, 041123 (2009).
\bibitem{Matsubara1987} Matsubara,~F. \& Inawashiro,~S. A frustrated antiferromagnetic Ising model on the hexagonal lattice. \textit{J.~Phys.~Soc.~Jpn.}~\textbf{56}, 2666--2674 (1987).
\bibitem{Mahendiran2003} Mahendiran,~R., Br\'eard,~Y., Hervieu,~M., Raveau,~B. \& Schiffer,~P. Giant frequency dependence of dynamic freezing in nanocrystalline ferromagnetic LaCo$_{0.5}$Mn$_{0.5}$O$_3$. \textit{Phys.~Rev.~B}~\textbf{68}, 104402 (2003).
\bibitem{Moessner2001} Moessner,~R., Sondhi,~S.~L. \& Chandra,~P. Phase diagram of the hexagonal lattice quantum dimer model. \textit{Phys.~Rev.~B}~\textbf{64}, 144416 (2001).
\bibitem{Basu2013} Basu,~T., Iyer,~K.~K., Singh,~K. \& Sampathkumaran,~E.~V. Novel dielectric anomalies due to spin-chains above and below N\'eel temperature in \cacoo. \textit{Sci.~Rep.}~\textbf{3}, 3104 (2013).
\bibitem{Bindu2009} Bindu,~R., Maiti,~K., Khalid,~S. \& Sampathkumaran,~E.~V. Structural link to precursor effects. \textit{Phys.~Rev.~B}~\textbf{79}, 094103 (2009).
\bibitem{Zorko2014} Zorko,~A., Adamopoulos,~O., Komelj,~M., Ar\v{c}on,~D. \& Lappas,~A. Frustration-induced nanometre-scale inhomogeneity in a triangular antiferromagnet. \textit{Nature Comms.}~\textbf{5}, 3222 (2014).
\bibitem{Ahn2004} Ahn,~K.~H., Lookman,~T. \& Bishop,~A.~R. Strain-induced metal-insulator phase coexistence in perovskite manganites. \textit{Nature}~\textbf{428}, 401–-404 (2004).
\end{thebibliography}
\end{document}